\documentclass[10pt, preprint2]{aastex}

\slugcomment{Ap. J., Submitted}

\shortauthors{Walker, Ohishi \& Mori}
\shorttitle{Microwave anisotropies}

\begin{document}

\title{Microwave anisotropies from the Galactic halo}

\author{Mark Walker}
\affil{School of Physics, University of Sydney, NSW 2006, Australia\\
\&\\
Australia Telescope National Facility, CSIRO\\}

\author{Michiko Ohishi and Masaki Mori}
\affil{Institute of Cosmic-Ray Research, University of Tokyo, Kashiwa, Chiba 277-8582, Japan}

\begin{abstract}
Models in which a large fraction of the Galactic dark
matter takes the form of cold gas clouds imply that there
is thermal microwave emission from the Galactic dark halo.
Such models can therefore be directly constrained by data
on the microwave sky, and in particular the very sensitive
observations of microwave anisotropies which are now being
made. To this end we have computed the anisotropy
power-spectrum expected for a Galactic dark halo made of
cold, dense gas clouds, including the effects of clustering
with a CDM-like mass spectrum of mini-halo substructure.
The power-spectrum displays two peaks: one, at $l\sim50$,
is the Poisson noise for the mini-halos, and the second,
much larger and at much higher $l$, is the Poisson noise
of the individual clouds. Because it appears on small
(milli-arcsecond) angular scales, where the instrumental
sensitivity is inevitably very poor, the latter signal
is not directly detectable. By contrast, clusters
of cold gas clouds may contribute significantly to the
observed anisotropies if their emission has a grey-body
spectrum. In this case the peak fluctuation, at $l\sim50$,
amounts to $4/|\sin b|\;{\rm\mu K}$ in the Rayleigh-Jeans
limit, and is the dominant Galactic foreground between
40 and 80~GHz. It will be possible to constrain this
foreground component using low-latitude data from the
MAP satellite, providing that its spectrum conforms to
a grey-body. If the spectrum is ``dusty'' there will be
relatively little power at frequencies below the thermal
peak, and in this case the predicted anisotropies are
shown to be negligible.
\end{abstract}



\section{Introduction}

Studies of the microwave sky play a critical r\^ole in
cosmology, and measurements of the degree-scale microwave
anisotropies can provide tight constraints on the
cosmological model (e.g. Hu and Dodelson 2002).
It is therefore particularly important that we understand the
power-spectrum arising from foreground microwave sources, and
much attention has already been given to this topic
(e.g. Tegmark and Efstathiou 1996; de~Oliveira-Costa and Tegmark 1999).
The various
foreground sources which have been considered to date include:
normal galaxies and Active Galactic Nuclei;
Galactic synchrotron emission;
Galactic free-free emission;
and Galactic dust emission. At frequencies in the range
$50\la\nu\la150$~GHz, none of these sources is thought to
make a significant contribution to the power-spectrum for
multipoles $l\la500$ (Tegmark and Efstathiou 1996).
This conclusion must be regarded as tentative in the case
of the extragalactic sources, because the difficulty of
undertaking large area surveys at high radio-frequencies
means that the source populations are poorly constrained
in this frequency range --- see Toffolatti et al (1999)
for a discussion. The Galactic
emission, on the other hand, is known to be very significant
at both low frequencies (synchrotron) and high frequencies
(dust); and, in fact, at all frequencies at low Galactic
latitudes. However, observing at high/low frequencies yields
template distributions for those contaminants, and these
templates can be cross-correlated with data in the preferred
frequency band to estimate their contamination level. This
procedure has provided robust confirmation that, at high
latitudes, the recognised Galactic foregrounds do not
significantly contaminate the COBE DMR data (Kogut et al 1996).

Although robust, the method we have just described clearly
cannot be applied if there is no template for the foreground.
This is the circumstance we are faced with if the Galactic
dark matter has a substantial component in the form of cold
gas clouds, as has been proposed by a number of authors
(Pfenniger, Combes and Martinet 1994; De~Paolis et~al 1995;
Henriksen and Widrow 1995; Gerhard and Silk 1996; Walker
and Wardle 1998; Sciama 2000a).
Such clouds emit primarily in the microwave band, and that
emission is thermal at temperatures of only a few degrees;
consequently the magnitude of this putative foreground is at
present only subject to very weak direct observational
constraints. Indirect constraints -- from the small amplitude
of fluctuations in the Cosmic Microwave Background (CMB), and
from Big Bang Nucleosynthesis  -- are generally supposed
to exclude any significant amount of baryonic dark matter
(e.g. Turner and Tyson 1999), but these arguments are not
entirely free of loopholes (Hogan 1993; Walker and Wardle
1999) and direct constraints are desirable.
Given the great sensitivity of the microwave data which
are now being acquired, it is therefore prudent to consider
what Galactic microwave emission is expected for models
in which the dark matter is composed of cold gas.

To proceed with a calculation
we need to specify a particular model. The success of
the now-standard structure formation paradigm,
exemplified by the Cold Dark Matter model
(e.g. Peebles 1992), argues that any acceptable
model must possess clustering properties that are
similar to those of CDM, in order to match the
data on large-scale-structure, and consequently we utilise
the properties of CDM halos as a guide. It then remains to
specify the properties of the individual clouds.
Radio-wave scintillation data -- specifically the
Extreme Scattering Events (Fiedler et~al 1987; Fiedler
et~al 1994) -- suggest a vast population of $\sim{\rm AU}$-sized
clouds, each of planetary mass (Walker and Wardle 1998),
and we assume that these properties are appropriate
to the individual clouds. 

The present paper is organised as follows:
in \S2 we derive the angular spectrum for an unclustered
cloud distribution; in \S3 the influence of CDM-like
clustering is considered; \S4 combines the results of
\S\S2,3 into a prediction for the root-mean-square
temperature anisotropy spectrum, under the assumption
of grey-body emission, and considers the constraints
which existing data place upon the input model; \S5
then considers how these results are modified in the
case where the clouds have ``dusty'' emission spectra.

\section{Power from individual clouds}
\label{sec:individual}
As a tool to describe the statistical properties of the
microwave sky, it is conventional to employ the power-spectrum
\begin{equation}
C_l\equiv{1\over{2l+1}}\sum_{m=-l}^{l}a_{lm}^*a_{lm},
\end{equation}
where the $a_{lm}$
are coefficients in the expansion of the brightness temperature
field, $\delta T$, in terms of the spherical harmonic functions $Y_{lm}$:
\begin{equation}
\delta T(\theta,\phi)=\sum_{l=0}^\infty\sum_{m=-l}^{l}a_{lm}Y_{lm}(\theta,\phi).
\end{equation}
We will model only the power-spectrum which would be
derived from measurements over a small patch of sky,
so that we need only consider the case $l\gg1$, for
which some mathematical simplifications are possible.
The restriction to large $l$ also means that we do
not have to model the structure of the Galaxy in any
detailed way, because we are considering angular
scales which are sufficiently small that the source
(cloud) positions are essentially random.

The starting point for our calculation is the white-noise
spectrum ($C_l=C_l^w$, independent of $l$) which is introduced
by a random distribution of point sources, each of flux $F$,
with a mean number per unit solid-angle of
${\rm d}N/{\rm d}\Omega$:
\begin{equation}
C_l^w = F^2 {{{\rm d}N}\over{{\rm d}\Omega}}.
\end{equation}
If the individual sources are not point-like, but
each source is axisymmetric with power-spectrum
$F^2C_l^s$ (so that $C_l^s$ is the power-spectrum for
a source of unit flux), then the power-spectrum of the
source population is $C_l=4\pi C_l^s\,C_l^w$. Now if we
consider sources at various distances, $D$, with
corresponding number density $n$, it follows that
$\delta({\rm d}N/{\rm d}\Omega)=D^2n\,\delta D$,
and hence
\begin{equation}
{{{\rm d}C_l}\over{{\rm d}D}}=4\pi C_l^s D^2 n F^2.
\end{equation}
Integrating along the line-of-sight then yields the
power-spectrum of this source population.

The gas clouds we are considering are expected to
cool primarily through thermal continuum emission
in the microwave band, with this emission arising
from precipitates of molecular hydrogen
(Wardle and Walker 1999). The bulk of the interior
of a cloud is too hot for solid hydrogen to exist;
all the radiation is expected to come from a thin,
nearly isothermal atmosphere. There is at present no
quantitative description of the internal cloud structure,
and consequently we have adopted a simple model of
the surface brightness profile, namely a uniform
intensity disk. Provided that the cloud radius is
$R_o\ll D$, this model yields a power-spectrum
\begin{equation}
C_l^s = {1\over{4\pi}}\left[{{2\,J_1(lR_o/D)}\over{lR_o/D}}\right]^2
\end{equation}
for a single cloud, with $J_1$ being a Bessel function
of the first kind. For $l\gg D/R_o$ this model is a
poor one in that it over-predicts
the power spectrum --- a result of the discontinuity
in the intensity profile at the limb of the cloud. To counteract
this tendency to overpredict the power at large $l$,
we have simply truncated the power spectrum at $lR_o=10D$.
This ad~hoc treatment of the high frequency behaviour
means that the predicted power-spectrum for the cloud
population is unreliable in the regime $lR_o\gg\langle
D\rangle$; however the corresponding angular scales are
so small (sub-milli-arcsecond; see \S4) that this
limitation is not important in practice.

In equilibrium the luminosity of any cloud must be
balanced by heating, so a cloud of mass $M_o$ satisfies
$\Gamma M_o = 4\pi D^2F$, where $\Gamma$ is the heating
rate per unit mass. This heating is expected to be due
primarily to cosmic-rays (Wardle and Walker 1999; Sciama
2000a). In this paper we consider only lines-of-sight at
high or mid Galactic latitude, and for these directions
it is reasonable to model the cosmic-ray density
variations with a simple analytic function; we employ
the form $\Gamma=\Gamma_0\exp(-|z|/h)$, where $z=D\sin b$
is the height above the Galactic disk at Galactic latitude
$b$. A more realistic model is offered by a sum of thin
and thick disks (Higdon 1979; Beuermann, Kanbach and
Berkhuijsen 1985), but for our illustrative calculations
this level of detail is unwarranted and we have adopted
a single component model. with a scale-height of $h=3$~kpc
(corresponding to the thick disk). This choice of
scale-height is a compromise between the (larger) values
deduced from analysis of the radio synchrotron data (Higdon
1979; Beuermann, Kanbach and Berkhuijsen 1985), and the
(smaller) values inferred from existing cosmic-ray
propagation models (Webber, Lee and Gupta 1992).
Note, however, that if the putative cloud population
does indeed exist, then these models of cosmic-ray
propagation need to be revised (Sciama 2000a, b).

To complete our prescription we need to specify the number
density of the clouds. For this purpose we employ the
collisional isothermal halo model of Walker (1999: W99). In this
model, or indeed any quasi-spherical halo model, the density
of dark matter varies only slowly in comparison to the scale-height
of the cosmic-ray disk, and this variation can be neglected:
\begin{equation}
n={\rho\over{M_o}}\simeq{{V^2}\over{4\pi GM_o(r_{gc}^2 + r_d^2)}},
\end{equation}
where $V\simeq220\;{\rm km\,s^{-1}}$ is the rotation speed
for the Galactic disk, $r_{gc}\simeq8.5$~kpc is the distance
to the Galactic Centre, and $r_d$ is the core radius of the
dark halo. For the Galaxy, the latter is specified as
$r_d\simeq6.2$~kpc in W99's model.

For a homogeneous source population, with $n\simeq$~constant,
and $F\propto 1/D^2$, equation 4 indicates that the low-frequency
power (for which $4\pi C_l^s\simeq1$)
is dominated by the nearest clouds within the
field-of-view, $\Delta\Omega$. The lower limit on the
integration over $D$ must therefore not be taken as zero,
but as $D_{min}$, the expected distance of the closest cloud, with
\begin{equation}
{{\Delta\Omega}\over{3}}D_{min}^3n:={1\over2}.
\end{equation}
For a field of
$\Delta\Omega\sim0.1$~sr, similar to the coverage of the
BOOMERanG experiment (de~Bernardis et al 2000), for example,
we have $D_{min}\simeq0.57$~pc.

\begin{figure*}
\plotone{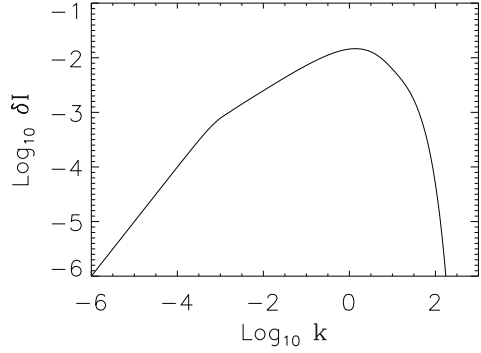}
\caption{The root-mean-square intensity fluctuations, $\delta I$,
due to a randomly distributed population of cold clouds, heated by
cosmic-rays, as described by equation 8. This graph is appropriate
for observations of a field of size $\Delta\Omega=0.1$~sr
near the Galactic pole ($|b|=90^\circ$). The axes are scaled
such that $k=2lR_o/h$, and $\delta I=k\sqrt{C_l^u(k)/C_l^u(0)}$.
\label{fig:unclustered}}
\end{figure*}

Using the descriptions given above, we can compute the
power-spectrum for our model of unclustered gas clouds, $C_l^u$:
\begin{equation}
{{C_l^u(k)}\over{C_l^u(0)}}=y\int_y^\infty\!\!{\rm d}x\,\exp(-x|\sin b|)
\left[{{2J_1(k/x)}\over{k}}\right]^2,
\end{equation}
where $x\equiv 2D/h$, so that $y=2D_{min}/h$, and $k\equiv2lR_o/h$.
The power level at zero frequency is 
\begin{equation}
C_l^u(0)={{M_oV^2\Gamma_0^2}\over{(4\pi)^3 G (r_{gc}^2 + r_d^2) D_{min}}}.
\end{equation}
Figure 1 shows a plot of $\delta I=k\sqrt{C_l^u(k)/C_l^u(0)}$,
the standard-deviation of the intensity
fluctuations. In constructing this figure we have taken
$|\sin b\,|=1$, and we have adopted
the parameters of the model given above, for which
$D_{min}\simeq0.57$~pc, hence $y\simeq3.8\times10^{-4}$.

Three distinct
regimes of behaviour can be seen in figure 1. For $k\ll y$,
all of the clouds are expected to be unresolved on the
corresponding angular scales, so the power-spectrum
is simply $C_l^u(k)\simeq C_l^u(0)$ and the r.m.s. intensity
fluctuations are proportional to $k$. For
$k\sim y$ some of the closest clouds are resolved out,
and for $y\ll k\ll1/|\sin b|$ the power flattens to
$C_l^u(k)/C_l^u(0)\simeq2y/\pi k$, so the r.m.s. intensity
is $\delta I\propto\sqrt{k}$. It should be noted that up
to this point the results are independent of the
scale-height, $h$, and the line-of-sight, i.e. the latitude $b$;
in other words the behaviour is completely determined
by the local conditions (dark matter density and cosmic-ray
heating rate). However, the location ($k=k_p\simeq\sqrt{2}/|\sin b|$)
and height of the peak fluctuations {\it are\/} influenced by $h$ and $b$:
clouds which lie outside the cosmic-ray disk have very low luminosity
and consequently contribute little to the temperature
anisotropies, so on angular scales smaller than those characteristic
of a cloud at distance $D\sim h/|\sin b|$ the fluctuations
decline rapidly. The peak value of the anisotropy spectrum
can be estimated from $C_l^u(k_p)/C_l^u(0)\simeq\sqrt{2}y|\sin b|/\pi$.
If we make numerical estimates of these values we find that
the peak occurs on milli-arcsecond scales ($l\sim3\times10^8$) and
reaches a value $\sim4$~mK, as discussed in \S4.

\subsection{Source counts}
In connecting the foregoing model with observations we
must take a little care, in that low-frequency power is
dominated by the brightest sources within the field-of-view.
If observations are made with a sensitive instrument that
has good angular resolution, then these will be recognised
as point sources and subtracted from the map before the
anisotropy power-spectrum is estimated. In order to quantify
this possibility we need to estimate the source counts.
For bright sources, with $D\ll h/|\sin b|$, the source
luminosity and density are uniform over the population,
hence the usual Euclidean result applies:
\begin{equation}
{{{\rm d}N(>F)}\over{{\rm d}\Omega}}={{1\over{4\pi}}}
\left({{F_{max}}\over{F}}\right)^{3/2},
\end{equation}
where $F_{max}$ is the expected flux of the brightest
source on the sky. For the case we are considering, the
bolometric luminosity is $\Gamma M_o$, so the brightest
source is expected to have a flux of
$F_{max}=\Gamma M_o/4\pi D_{min}^2$.
This result breaks down at low flux levels,
where the sources lie outside the cosmic-ray disk, but
$F_{max}$ corresponds to a source which is only about
0.1~pc distant, so with a disk scale-height of $h=3$~kpc
the form given in equation 10 is a good approximation
for $F\gg10^{-9}F_{max}$. Numerical estimates for $F_{max}$
are given in \S4.

\begin{figure*}
\plotone{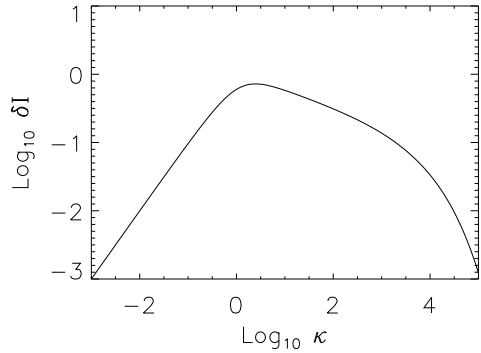}
\caption{The root-mean-square intensity fluctuations,
$\delta I=\kappa\sqrt{C_l^c(\kappa)/C_l^c(0)}$, due to a
population of mini-halos with a CDM-like mass spectrum.
This plot is appropriate to a survey region of size
$\Delta\Omega=0.1\;{\rm sr}$ near the Galactic pole.
At low frequencies ($\kappa\ll1$) essentially all of
the mini-halos are unresolved and $\delta I\simeq\kappa$,
reaching a peak of $\delta I\simeq1$ at $\kappa\simeq1$.
This peak reflects the characteristic angular size of
the closest mini-halos, which is of order $4^\circ$.
\label{fig:clustered}}
\end{figure*}

\subsection{Mean sky brightness}
Using the foregoing model, it is straightforward to
calculate the mean sky brightness contributed by the
cloud population, the result is:
\begin{equation}
\langle I\rangle={{\Gamma_0V^2h\;{\rm csc}|b|}\over
{(4\pi)^2G(r_{gc}^2+r_d^2)}}.
\end{equation}
The mean sky brightness thus does not depend
explicitly on the properties of the individual clouds,
although there is an implicit dependence of $\Gamma_0$ on
the column-density of the clouds, as we shall see
in \S4.1. Consequently the same result for $\langle I\rangle$
is obtained for the model described in the next section,
in which clustering of the clouds is considered.
Using the cosmic-ray heating rate determined in \S4.1, and an
atmospheric temperature of 4.2~K for the clouds (\S4.2),
we deduce a mean sky brightness temperature, in the
Rayleigh-Jeans regime, of
$\langle T_b\rangle\simeq30\;{\rm\mu K}$
at the Galactic pole.

\section{Power due to clustering}
\label{sec:clustering}
The Cold Dark Matter model provides an elegant framework
within which we can understand the formation of structure in
the Universe (Blumenthal et al 1984; Davis et al 1985;
Peebles 1992). The key ingredients of this model are (i) a
scale-invariant power-spectrum of density fluctuations,
and (ii) dark matter which initially has low velocities.
Gravity does the rest. Dense gas clouds are expected to
satisfy the second of these requirements (i.e. they are
a type of cold dark matter), provided that they form in
the early Universe and survive to the present time (Hogan
1993; Walker and Wardle 1999). The first requirement is
naturally satisfied for inflationary models of the early
Universe, independent of the nature of the dark matter.
Consequently structures forming in a Universe dominated
by dense gas clouds may resemble those found in CDM models.
This is as true for galactic and sub-galactic mass-scales
as it is for cluster and super-cluster scales. The
substructure within CDM halos follows a power-law mass
spectrum with index approximately equal to $-2$ (Moore
et al 1999; Klypin et al 1999), and we adopt this result
as our model for the clustering properties of dense gas
clouds comprising the Galactic dark halo. Consequently
we expect the halo to be populated with a number density
$n_{mh}$ of mini-halos, such that
\begin{equation}
{{{\rm d}n_{mh}}\over{{\rm d}M}}={{\rho}\over{M^2\log_e(M_2/M_1)}}
\end{equation}
extending over a mass range $M_1<M<M_2$. It should be
noted that this formulation implies that {\it all\/} of the
dark matter is in the form of mini-halos, and this is
apparently at odds with the results of simulations,
for which only $\sim10$\% of the halo mass appears
in this form (e.g. Ghigna et al 1998). However, the simulations
which have been undertaken to date are capable of resolving
only the upper end of the mini-halo mass range which we
consider here (see \S4.3), and such simulations have
little to say about the existence of lower mass halos.
Moreover, if the only relevant physics is cold
gravitational collapse, acting on a power-law spectrum
of density perturbations, then there is no reason to
expect any characteristic mass-scale in the clustering,
until one reaches masses which are so small that the
discreteness of the dark matter particles begins to play
a role. At the very least the circumstance described by
equation (12) can be considered as a limiting case: the
total density in mini-halos cannot exceed the total dark
matter density.

Because these aggregates are much smaller than their
parent halos, their internal structure has not yet been
thoroughly investigated with simulations. However, we
know that their density profiles are consistent with
tidally-truncated isothermal spheres (Moore et al 1999),
and we adopt that description. In addition to tidal
truncation, the density distribution must have a central
core as a result of collisions occurring between the
constituent clouds (W99). We therefore employ a model
density profile
\begin{equation}
\varrho={{M\exp(-r^2/2r_t^2)}\over{(2\pi)^{3/2}r_t(r^2+r_c^2)}},
\end{equation}
where $r_t$ is the tidal radius and $r_c$ the core radius
of the mini-halo, and we have assumed that $r_c\ll r_t$.
From this three-dimensional profile we can determine
the column-density seen by a distant observer. The microwave
intensity is proportional to this column, so 
the angular power-spectrum for a source of unit flux is
\begin{equation}
C_l^s\simeq{{\exp(-2lr_c/D)}\over{4\pi + 8l^2r_t^2/D^2}}.
\end{equation}
The features of this power-spectrum are as follows. For
$lr_t\ll D$, $C_l^s\simeq1/4\pi$; and for $D/r_t\ll l\ll D/r_c$
we have $C_l\propto 1/l^2$, cutting off exponentially
for $l\ga D/r_c$. Most of the contribution to the
intensity fluctuations therefore comes from the
frequency range $D/r_t\ll l\ll D/r_c$, and the existence of
a core in the density profile is critical in determining
the power-spectrum at high wave-numbers. If there were
no core in the halo density profile, the intensity
fluctuations would be independent of $l$ in the range $l\gg D/r_t$.

We need to calculate the tidal radius and the
core radius; both depend on the
mass of the mini-halo under consideration. To estimate
the tidal radius we note that the scale-height of
the cosmic-ray disk,
$h$, is expected to be small in comparison with
$r_{gc}$, so we can make the approximation that
all of the mini-halos within the field-of-view which
make a significant contribution to the power-spectrum
are at the same Galactocentric radius $r=r_{gc}$.
Now the tidal radius of each mini-halo is determined
by the tides at perigalacticon, and because $r_{gc}$ is
an upper limit to the galactocentric radius at that
point -- i.e. most mini-halos will be on orbits which
take them much closer to the centre of the Galaxy
-- and simulations of halo formation find that
mini-halo orbits have quite large eccentricities
(Ghigna et al 1998),
we take $r_{gc}/2$ as an estimate of the Galactocentric
radius at perigalacticon. We then follow the usual
procedure (Binney and Tremaine 1987), leading to an
estimated tidal radius $r_t^3:=\alpha M$, where
\begin{equation}
\alpha={{Gr_{gc}^2}\over{12V^2}}
\simeq8.0\times10^{21}\;{\rm cm^3\,g^{-1}}.
\end{equation}
The core radius of each halo is determined by the collisional
nature of the dark matter (W99), with the
velocity dispersion of each mini-halo being related to
its mass by comparing equation 13 (for $r\ll r_t$)
with the standard
form for a self-gravitating isothermal sphere. This
procedure leads to the estimate $r_c^2=\gamma M$, where
\begin{equation}
\gamma = {{{12\sqrt{2}\, Vt_H}\over{(2\pi)^{3/4}\,\pi\Sigma r_{gc}}}}
\simeq2.5\;{\rm cm^2\,g^{-1}},
\end{equation}
with $t_H\simeq10^{10}$~years being the elapsed time since
the mini-halo virialised, and $\Sigma\equiv M_o/\pi R_o^2$
is the mean column-density of the individual clouds.
W99's model gives $\Sigma\simeq140\;{\rm g\,cm^{-2}}$,
and we have used that value in estimating $\gamma$.

The power-spectrum for the whole population of mini-halos
can be obtained by integrating  
\begin{equation}
{{{\rm d^2}C_l}\over{{\rm d}D\,{\rm d}M}}=
4\pi C_l^s D^2 {{{\rm d}n_{mh}}\over{{\rm d}M}} F^2,
\end{equation}
with respect to $M$ and $D$. The integration over distance
is taken out to infinity but, as for the case of the individual
clouds, the lower limit cannot be taken as zero because
at low frequencies the
power-spectrum is dominated by contributions from the nearest
sources. We can estimate the lower distance limit, as a
function of mini-halo mass, by analogy with equation 7,
making the replacement
$n\mapsto\log_e(M_2/M_1){{\rm d}n_{mh}}/{{\rm d}\log_eM}$,
whence
\begin{equation}
{{\Delta\Omega}\over{3}}D_{min}^3 \log_e(M_2/M_1)
{{{\rm d}n_{mh}}\over{{\rm d}\log_eM}}={1\over2}.
\end{equation}
This equation states that the closest mini-halo
within the mass range $M_1$ to $M_2$ has a 50\% chance
of being located within a distance $D_{min}$.
If we now define $z\equiv2D_{min}/\tilde{h}$, where
$\tilde{h}\equiv h/|\sin b|$ then with reference to
the expected mini-halo mass spectrum (equation 12) we can
write $z^3:=\beta M$, where
\begin{equation}
\beta = {{12}\over{{\tilde{h}}^3\, \Delta\Omega\,\rho}}
\simeq2.7\times10^{-40}\;{\rm g^{-1}}.
\end{equation}
In estimating the numerical value of $\beta$, here, we have
assumed a field-of-view of $\Delta\Omega=0.1\;{\rm sr}$
at the Galactic pole ($|\sin b\,|=1$), with the mass limits
$M_{1,2}$ as given in \S4.
An important point can now be made, even before we calculate
the power-spectrum of the population. Because the tidal
radius, $r_t$, and the distance to the closest mini-halo,
$D_{min}$, both vary as $M^{1/3}$, the angular radius of
the nearest mini-halo is
$r_t/D_{min}=(2/\tilde{h})(\alpha/\beta)^{1/3}\sim4^\circ$,
{\it independent of $M$,\/} (and also independent of $h$
and $b$).
Remembering that the power-spectrum is
dominated by contributions from the closest mini-halos,
we thus expect a peak in the intensity fluctuations on
degree scales.

We can now integrate equation 17 over $D$ and $M$ to
obtain our power-spectrum estimate; the result is
\begin{equation}
C_l^c=C_l^c(0)\,{\cal I}(\kappa,\kappa_c),
\end{equation}
where
\begin{equation}
C_l^c(0)={{\Delta\Omega}\over{6\log_e(M_2/M_1)}}\left({{V}\over{4\pi}}\right)^4
\left[{{\Gamma_0\tilde{h}}\over{G(r_{gc}^2 + r_d^2)}}\right]^2,
\end{equation}
and
\begin{equation}
{\cal I}(\kappa,\kappa_c)=3\int\!{\rm d}z\,z^2\!\!\int_z^\infty
\!\!\!\!\!{\rm d}x\,{{\exp(-x-\kappa z^{3/2}/\kappa_cx)}\over{x^2+\kappa^2z^2}},
\end{equation}
having introduced $x\equiv 2D/\tilde{h}$.
The wavenumber is now expressed as
$\kappa\equiv(l/\tilde{h})(\alpha/\beta)^{1/3}(8/\pi)^{1/2}$,
and the parameter $\kappa_c$ is
$\kappa_c:=\alpha^{1/3}\beta^{1/6}/\sqrt{2\pi\gamma}$.
The limits of integration
over $z$ are determined by the limits of the mini-halo
mass-spectrum, so $z_{1,2}^3:=\beta M_{1,2}$, and
we have adopted the values of $M_{1,2}$ as estimated
in \S4, leading to $z_1=3.8\times10^{-3}$ and $z_2=9.1$.

It is easy to verify that ${\cal I}\rightarrow1$ as $\kappa\rightarrow0$.
Unfortunately this low frequency regime (i.e. $\kappa\ll1$) is the
only one for which we have been able to determine a simple
analytic form for ${\cal I}$. It is, however, straightforward
to evaluate ${\cal I}$ numerically, once $\kappa_c$ is
specified. Using the numerical values for the various parameters
which we have already given (see also the following section),
we find $\kappa_c\simeq1.3$, leading to the fluctuation
spectrum shown in figure 2. At low frequencies, where essentially
all of the mini-halos are unresolved, this spectrum
displays the usual form for Poisson noise from
point-like sources: $\delta I\propto\kappa$. The spectrum
then peaks at $\kappa\sim1$, corresponding to the
characteristic angular scale of the nearest mini-halos,
and declines slowly as $\kappa$ increases.

\section{Model parameters}
\label{sec:model}
The anisotropy models we have just presented do not
incorporate any free parameters: all of the various
inputs can be estimated -- albeit rather crudely in
some cases -- from other types of data. We now turn
to the business of estimating those quantities which
have not yet been specified, in order to arrive at
numerical results for the brightness-temperature
anisotropy spectrum.

\begin{figure*}
\plotone{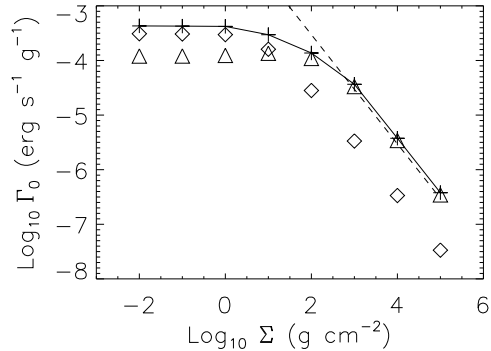}
\caption{The cosmic-ray heating rate for dense
gas clouds, as a function of the column-density
of the cloud. The points are the results of computations
with the GEANT Monte Carlo simulation package: diamonds
show the electron contribution (primary cosmic-ray
electrons of energy greater than 10~MeV); and
triangles show the proton contribution (protons of
energy greater than 100~MeV). The sum of these two
components gives us the total heating rate shown by
the crosses (connected with a solid line). The dashed
line shows the limiting case, appropriate to very
dense gas clouds, in which all of the incident cosmic-ray
power is absorbed. The fact that the GEANT results
lie slightly above this line, for high gas columns,
is a reflection of energy non-conservation in
hadronic interactions within GEANT.
\label{fig:heating}}
\end{figure*}

\subsection{Cosmic-ray heating}
For high column-density clouds, such as we are considering
here ($\Sigma\simeq140\;{\rm g\,cm^{-2}}$, W99),
there is significant attenuation of the cosmic-ray flux
from surface to centre, and so the cosmic-ray heating rate
differs from that deduced for diffuse interstellar gas
(Umebayashi and Nakano 1981; Sciama 2000a).
We have estimated the cosmic-ray heating
rate local to the Sun, $\Gamma_0$, appropriate to dense
clouds, using the Monte Carlo simulation program GEANT
(Ohishi, Mori and Walker 2003; {\tt http://wwwinfo.cern.ch/asd/geant}).
GEANT was developed for studying interactions in laboratory
particle detectors, but can be used for astrophysical
calculations without modification. By using GEANT we are able
to study the transport of cosmic rays into high column-density
clouds, with all relevant particle interactions
included. The main disadvantage of GEANT, for our
purposes, is that it does not conserve energy
very accurately in the hadronic interactions
(GEANT Bug Reports \#171 \& \#389). There is a ``patch''
available for the relevant piece of code, GHEISHA which
treats hadronic interactions, but
this patch appears to introduce further problems of its
own (GEANT Bug Report \#415), and we have therefore
used GEANT with the un-patched version of GHEISHA.
We can easily calculate the heating rate in the limiting
case of very dense clouds, without using GEANT, because
all of the incident power is absorbed in this case. We
find that in this limit GEANT over-estimates the
power deposition by $\sim$25\%, and this estimate
can be taken as a gauge of the accuracy of
our computations of the heating rate, $\Gamma$.

For the purposes of estimating the microwave emissivity
of the gas, we need only calculate the total power deposited
by the incident cosmic-rays; however, the interaction between
the cosmic-rays and the gas also gives rise to gamma-rays
(principally via electron bremmstrahlung and ${\rm\pi^o}$ production),
and we will report elsewhere on the consequences of these
processes for the interpretation of the mean gamma-ray
spectrum of diffuse Galactic emission (Ohishi, Mori and Walker 2003),
and for the interpretation of the unidentified EGRET sources
(Walker, Mori and Ohishi 2003).  In our calculations of the
gas heating rate we approximated each cloud as a sphere of
uniform density, and used GEANT to compute the power
deposited as a function of the gas density (the radius of
each cloud being fixed at 1~AU). We adopted
the ``median'' cosmic-ray proton spectrum quoted by Mori (1997;
but note that the units on his equation [3] should read
${\rm cm^{-2}s^{-1}sr^{-1}GeV^{-1}}$),
including energies down to 100~MeV, and the electron
spectrum from Skibo and Ramaty (1993),
with a low-energy cutoff of 10~MeV; more
details of these simulations will be presented elsewhere
(Ohishi, Mori and Walker 2003).
These cosmic-ray spectra are appropriate to the vicinity
of the Sun, and it is implicit in our model that the spectra
elsewhere in the Galaxy have the same shape, and differ only
in normalisation. We caution that the cosmic-ray spectra
which we are employing are not identical to the spectra which
are directly observed, but are ``demodulated'', to correct
for the influence of the solar magnetosphere. At low
particle energies the demodulation corrections are
large, with large associated uncertainties. 

Our results are shown in figure 3, in the
form of heating rate per unit mass, as a function of the
mean column-density, i.e. $\Gamma_0(\Sigma)$.
Electron and proton contributions
are shown separately (points) and in combination (solid line).
Taken separately the electrons and protons display the same
qualitative behaviour: at low columns the specific heating rate
is approximately constant, while at high columns it declines
according to $\Gamma_0\propto\Sigma^{-1}$. For low column-densities
our calculations yield cosmic-ray heating rates of order
$3\times10^{-4}\;{\rm erg\,s^{-1}\,g^{-1}}$ which are
consistent with previously reported cosmic-ray ionisation
rates appropriate to diffuse interstellar gas (e.g. Webber 1998),
given an absorbed energy of approximately 7~eV per ionisation
(Glassgold and Langer 1973; Cravens and Dalgarno 1978).
However, we caution that our calculated heating rates in
the low column-density regime are sensitive to the low-energy
cutoff of the primary cosmic-ray electrons included in
the simulation. More generally, this finding implies that
at low column-density the heating rate is sensitive to
the form of the adopted cosmic-ray electron spectrum at
low energies, and in this regime the spectrum is poorly
constrained. The uncertainties in cosmic-ray spectra have been
highlighted by the recent determination (McCall et al 2003)
of an ionisation rate, in a diffuse cloud (column
$\sim3\times10^{-3}\;{\rm g\,cm^{-2}}$), which is 40
times higher than previously assumed. This suggests
that $\Gamma_0$ may be as large as
$10^{-2}\;{\rm erg\,s^{-1}\,g^{-1}}$ at the extreme
left of figure 3. However, for the column densities
which concern us here the heating rate is determined
primarily by the total energy density in cosmic-rays
(see below), and it is not clear whether the new
data imply any significant revision of this quantity.

The behaviour of the heating rate at high/low columns is
straightforward to understand: for small
columns the cloud is ``optically thin'' to the cosmic rays,
so that the absorbed power is directly proportional to the
mass of the cloud and the specific heating rate is
independent of the column. At
large columns essentially all of the power incident on the
cloud surface is absorbed, so that the total heating
rate is proportional to surface area, independent
of the cloud mass, and the specific heating rate
is therefore inversely proportional to the column-density.
In this regime the computed heating rate is not
sensitive to the details of the cosmic-ray spectrum, being
dependent only on the total energy density resident in
the cosmic rays.
We are primarily interested in high column-density gas,
for which the heating is mainly due to protons --- most
of the cosmic-ray energy-density being resident in protons
of energy $\sim\,$few~GeV. For our preferred column-density of
$\Sigma\simeq140\;{\rm g\,cm^{-2}}$ (see \S4.2), the heating rate
is $\Gamma_0\simeq10^{-4}\;{\rm erg\,s^{-1}\,g^{-1}}$;
80\% of this power is contributed by the protons.

\subsection{Properties of individual clouds}
Individually the cloud mass and radius are poorly known,
but the mean column-density, $\Sigma\equiv M_o/\pi R_o^2$,
is quite tightly constrained because it determines
the rate of conversion of dark to visible matter
(W99). Matching the model
to the Tully-Fisher relation for late-type spirals yields 
$\Sigma\simeq140\;{\rm g\,cm^{-2}}$ (W99).
Essentially all of the uncertainty is then
in the choice of cloud mass. The viable mass range
is currently rather large, perhaps as broad as
$10^{-6}\,{\rm M_\odot}\la M_o\la10^{-2}\,{\rm M_\odot}$
(Wardle and Walker 1999). At present the best estimate
we can make for the cloud mass comes from the optical
variability of quasars: Hawkins (1993, 1996) pointed out that
the observed optical variations, on a time-scale of years,
could be interpreted in terms of a near-critical Universal
density of planetary-mass gravitational lenses. In the
context of our model, these lenses are to be identified
with the individual gas clouds (whereas Hawkins[1993]
suggested that they might be primordial black holes).
Hawkins's (1993) estimate of the mass of the individual
lenses is $10^{-3}\,{\rm M_\odot}$. Schneider (1993),
however, has modelled the same data set and his analysis
suggests a mass of $10^{-4}\,{\rm M_\odot}$ if the
observed variability is interpreted as lensing. Minty
(2001) re-examined this issue and demonstrated consistency
between lensing models and the data for lens masses
in the range $10^{-5}-10^{-4}\,{\rm M_\odot}$. The lower
end of this range is, however, excluded by data on
Galactic microlensing (Rafikov and Draine 2001), because
of refraction in the neutral gas (Draine 1998). We adopt
the value $10^{-4}\,{\rm M_\odot}$. This choice then
determines the cloud radius, because the column-density
is fixed; correspondingly $R_o\simeq1.4\;{\rm AU}$, close
to our adopted value (\S4.1) of 1~AU.

\begin{figure*}
\plotone{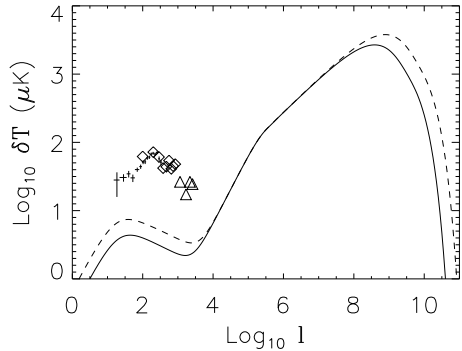}
\caption{Microwave anisotropy spectra due to emission from
cold, Galactic clouds. This plot shows the root-mean-square
brightness-temperature fluctuations,
$\delta T=[l(l+1)(C_l^u + C_l^c)/2\pi]^{1/2}$, as a
function of spherical harmonic index, $l$, in the Rayleigh-Jeans
limit. The large peak at $l\sim3\times10^8$ is the Poisson noise
from individual clouds, while the smaller peak at lower $l$ is due
to bound clusters of clouds (mini-halos) with an assumed CDM-like
mass spectrum. The figure shows the predicted spectrum appropriate
to a small area of sky ($\Delta\Omega\sim0.1\;{\rm sr}$) at high/low
Galactic latitude (solid/dashed lines: $|b|=90^\circ$/$|b|=30^\circ$
respectively), and data taken at high latitude with the DASI instrument
(diamonds; Pryke et al 2002), and the CBI (triangles; Mason et al 2002).
Also shown on this panel are the results from Archeops (crosses;
Beno\^\i t et al 2002), based on data covering an area
$\Delta\Omega\simeq1\;{\rm sr}$ above $b=30^\circ$.
\label{fig:both}}
\end{figure*}

The other key aspect of the individual clouds is their
emission spectrum. The model of Wardle and Walker (1999),
which focusses on the thermal stability of the clouds,
requires that they be cooled by thermal continuum emission
from  ${\rm H_2}$ precipitates. Here we assume, initially,
that the hydrogen ``snowflakes'' are large compared to
the $\sim\,$mm wavelengths of their thermal emission,
and that the individual snowflakes have a large optical
depth to absorption at these wavelengths. These assumptions
imply grey-body emission ($I_\nu\propto B_\nu$) at the
temperature of the ``snowflake''. (Later, in \S5, we will
examine how the model predictions are modified if we
relax these assumptions.) The latter
is, to a very good approximation, just the temperature of
the atmosphere of the cloud. We noted earlier that
the interior of the cloud is likely to be convective,
and we can therefore make an estimate of the the
atmospheric temperature by calculating where an ${\rm n}=3/2$
polytropic model crosses the phase-equilibrium curve
for hydrogen. Fixing the mean column density at
$\Sigma=140\;{\rm g\,cm^{-2}}$, this method yields the
following estimated atmospheric temperatures:
$5.8,\;4.9,\;4.2,\;3.7,\;3.4$~K, for clouds of mass
$10^{-6},\,10^{-5},\,10^{-4},\,10^{-3},\,10^{-2}\;{\rm M_\odot}$,
respectively. For our preferred cloud mass, then, the
atmospheric temperature is roughly 4.2~K, and this
corresponds to a conversion factor of
$\delta T/\delta I=752\;{\rm K\,erg^{-1}\,s\,cm^2\,sr}$
between brightness-temperature, in the Rayleigh-Jeans limit,
and bolometric intensity. In the formulation we have
given, the root-mean-square intensity spectrum
-- i.e. the quantity $\delta I=\sqrt{l(l+1)\,C_l/2\pi}$ -- has
the units of bolometric intensity, and must be multiplied
by this conversion factor to deduce the temperature anisotropy.

Similarly, when considering source counts, the conversion
factor between flux density and bolometric flux for our
model parameters is
$F_\nu/F=2.3\times10^{-16}\nu^2\;{\rm Hz^{-1}}$,
in the Rayleigh-Jeans regime, with $\nu$ in GHz.
Using these conversion factors we can estimate the
flux of the closest single cloud, it is
$F_{max}\simeq10\;{\rm nK\,sr}$, implying that it
may just be detectable by the MAP satellite
({\tt http://map.gsfc.nasa.gov}). Sciama (2000a)
has previously noted that the proposed Planck
satellite ({\tt http://astro.estec.esa.nl/Planck})
would easily detect the closest examples of
the posited clouds. In more conventional units the
expected flux of the brightest source can be writen
as $F_{max}\simeq120\;{\rm mJy}$ at 20~GHz, and
scaling as $\nu^2$ in the Rayleigh-Jeans regime
($\nu\ll150$~GHz).

\subsection{Properties of the minihalos}
The mass-spectrum of the mini-halos is given in equation 12,
but we have not yet specified the range $[M_1, M_2]$
over which it applies. It is implicit in equation 12 that
the (dark) density $\rho$ is entirely composed of mini-halos
with masses in the range $M_1<M<M_2$. The appropriate value
of $M_1$ is difficult to estimate. To form a stable cluster
there should be a large number of clouds in each mini-halo,
although the choice of what constitutes a large number
is somewhat arbitrary; we have chosen
$M_1=0.1\;{\rm M_\odot}\sim10^3 M_o$.
The upper limit is easier to estimate: $M_2$ should be
chosen such that there is a 50\% chance of finding
a more massive halo within the field-of-view. If the total
mass of the dark halo of our Galaxy is
$M_{tot}$, which we take to be $2\times10^{12}\;{\rm M_\odot}$
(Zaritsky 1999), then it follows that
\begin{equation}
{{1\over2}}\sim{{\Delta\Omega}\over{4\pi}}
{{M_{tot}}\over{M_2 \log_e(M_2/M_1)}},
\end{equation}
and solving yields $M_2\simeq1.4\times10^9\;{\rm M_\odot}$
for $\Delta\Omega=0.1\;{\rm sr}$.
Fortunately our final results are not very sensitive to the
particular values of $M_{1,2}$ adopted, as they enter principally
through the factor $\log_e(M_2/M_1)\simeq23$, and secondarily
through the limits of integration ($z_{1,2}\propto M_{1,2}^{1/3}$).

\subsection{Temperature anisotropies and constraints}
With the above numerical estimates, and the power-spectrum
formulations derived in \S\S2,3, we are now in a position to
quantify the temperature anisotropies expected in the present
model. The results are graphed in figure 4, showing both the
mini-halo contribution, which leads to the peak at $l\sim50$,
and the much larger peak at $l\sim3\times10^8$ is the Poisson
noise from individual clouds.

Recalling, from \S2.2, that the mean sky brightness
is approximately $30\;{\rm \mu K}$, in our model, we see
that the mini-halos
introduce fluctuations which are small in comparison with
the mean intensity, while the reverse is true for the
fluctuations due to the individual clouds. This difference
simply reflects the fact that the mini-halos are sufficiently
large that they cover the entire sky several times over, whereas
the individual clouds cover only a tiny fraction of the
sky and the root-mean-square intensity is consequently
much greater than the mean. It is worth emphasising that
each of these contributions is computed under the assumption
that {\it all\/} of the dark matter is in the corresponding
form -- i.e. clustered into mini-halos, or entirely unclustered
-- and the calculated contributions are in this sense mutually
exclusive. However, recognising that (i) the minihalos are
made up of individual clouds, (ii) the minihalos
are unbiased tracers of the mean density, and (iii) the
minihalos introduce only a small modulation around
the mean sky brightness, we see that in practice
the predicted properties of the $\sim$milli-arcsecond
fluctuation peak are largely independent of whether
or not the clouds are clustered into minihalos.
If all the clouds are clustered into minihalos, then
there would, however, be many fewer clouds very local
to the Sun, and so the low-frequency (i.e. small $l$)
power from individual clouds would be much reduced.

How do our predictions compare with existing data?
We do not consider, here, the low-order multipoles
($l\la10$) which were measured by COBE (Smoot et al 1992),
because the calculations we have undertaken are valid only
for $l\gg1$. The COBE detections at small $l$ did, however,
create great interest in measuring the higher order
multipoles, and after much effort expended on special-purpose
experiments there are now several clear detections of
anisotropies on degree scales (e.g. de~Bernardis et al 2000;
Hanany et al 2000; Pryke et~al 2002; Beno\^\i t et al 2002).
The peak of the observed signal -- $\delta T_b\simeq70\;
{\rm\mu K}$ at $l\simeq200$ -- is roughly 20 times larger
than the model predictions at the same angular scale, for
high-latitude fields, and for many purposes this small
contribution to the observed power (about 0.3\%) can be
neglected. However, it may still be possible to detect
the predicted mini-halo anisotropies against the background
of the dominant CMB fluctuations, if we select the
data appropriately. In particular the CMB anisotropies
fall to $\simeq30\;{\rm\mu K}$ at multipoles $l\sim50$,
near the peak of the predicted spectrum, while the
latter attains a value $\simeq8\;{\rm\mu K}$ at 
$|b|=30^\circ$. Data collected by the MAP
satellite\footnote{\tt http://map.gsfc.nasa.gov}
should be able to reveal this Galactic foreground,
provided that the spectrum is close to the assumed
grey-body form (see \S5 and figure 5).

\begin{figure*}
\plotone{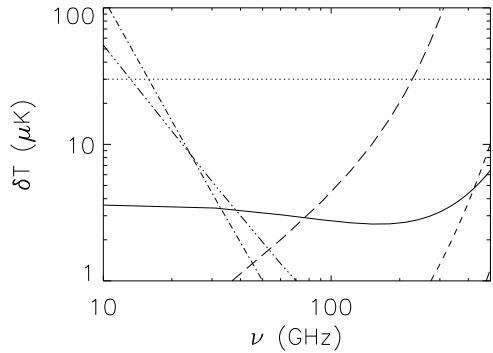}
\caption{Frequency dependence of the root-mean-square
temperature anisotropies at Galactic latitude $|b|=90^\circ$
and multipole $l\simeq50$, for various processes. The curves
shown here are for: the microwave background anisotropies
(dotted line; $T=2.73$~K, and $\delta T_b=30\;{\rm\mu K}$);
mini-halo anisotropies for a grey-body spectrum at $T=4.2$~K
(solid line); mini-halo anisotropies for a ``dusty'' spectrum
(i.e. $I_\nu\propto\nu^4$ at low frequencies) at $T=4.9$~K
(short-dashed line); emission from dust ($T=21$~K) in the
diffuse interstellar medium (long-dashed line); synchrotron
emission (dot-dashed line); and free-free emission
(dot-dot-dot-dashed line). The last three of these are
taken from Tegmark and Efstathiou (1996).
\label{fig:temp}}
\end{figure*}

To date there has been only limited interest in the
arcminute-scale anisotropies, because Silk damping
is expected to strongly supress any primary cosmological
anisotropy on these scales (Silk 1968). Consequently there
is at present only a limited amount of data in the
region $10^3\la l\la10^4$ (Subrahmanyan et al 1997;
Dawson et al 2001; Mason et al 2002). Here again the
observed signal is large in comparison with the model:
$25\;{\rm\mu K}$ at $l\sim2500$ (Mason et al 2002);
roughly an order of magnitude larger than the anisotropies
predicted by the present model. Consequently data at
these very high multipoles do not yet provide strong
constraints on the model we have presented. Moreover
these angular scales are below the resolution of the
instrumentation carried by MAP, so there is no immediate
prospect of a major improvement in sensitivity. It would,
however, be useful to obtain further measurements of the
power-spectrum using ground-based interferometers,
particularly with a view to constraining its latitude
dependence, because the origin of the observed
high-frequency power (Mason et al 2002) is not clear
at present and it remains possible that it is a
Galactic signal.

For $l\gg10^4$ the anisotropies predicted
by our model are dominated by the individual clouds,
and more specifically the closest examples within the
area under study; consequently for these angular scales
the model predicts very little latitude dependence in
the power-spectrum. The domain of large $l$ is also
technically challenging, in that the sensitivity of
an interferometer to surface brightness fluctuations
worsens in proportion to $l$.
Thus the predicted milli-Kelvin
peak on milli-arcsecond scales is not immediately
open to experimental scrutiny. Fortunately the model
predictions can be put to the test in another way:
rather than attempting to measure the anisotropies
per se we can simply study the bright source counts.
At low frequencies the flux from each cloud increases
as $\nu^2$, or faster, whereas the flux from non-thermal
(synchrotron) radio sources typically declines with
increasing frequency. Thus in a sky-survey for compact
sources at high frequencies the predicted population
should stand out. For the purposes of testing the model,
the ideal approach is to make a survey of
a large fraction of the sky, with sufficient sensitivity
to detect sources having flux much less than $F_{max}$.
The model then predicts that many sources with thermal
spectra should be detected, and any such sources can
be selected out and subjected
to detailed studies to determine their nature.

To date surveys at high radio frequencies have covered
only a very small fraction of the sky with the necessary
sensitivity --- this is simply a consequence of the
rapid decrease in area of the primary beam of an antenna,
as the observing frequency increases, requiring a much
larger number of pointings to cover a given area of sky.
For example, the survey of Taylor et al (2001), made with
the Ryle Telescope, covered only $63\;{\rm deg^2}$ at
$\nu=15$~GHz, to a flux limit of 20~mJy. Using the
predicted source-counts for our model (\S\S2.1,4.2), it
is straightforward to estimate that the expected number
of cloud detections for this survey is approximately
0.01 and these data therefore have little to say about
the present model. A more extensive interferometric
survey, covering the entire southern sky, is planned
for later this year. This survey is expected to have a
flux limit of roughly 40~mJy at 20~GHz -- thus improving
substantially on the pilot survey (Ekers et al 2003)
-- and could detect examples of the closest clouds
if their spectra are indeed grey bodies (see \S5).

Deep surveys at much higher frequencies have also been
undertaken, again covering very small areas of the
sky. Interestingly, Lawrence (2001) has pointed out
that the ``blank field SCUBA sources'' are consistent
with being a very local population of Galactic sources,
rather than high-redshift star-forming galaxies as
is usually assumed. Indeed Lawrence (2001) points out
that their properties are remarkably similar to those
expected for the population of cold gas clouds postulated
by Walker and Wardle (1998); the main point of discrepancy
noted by Lawrence is that he infers a thin disk
population, rather than a quasi-spherical halo. However,
as noted earlier, the luminosity of each cloud is determined
by the cosmic-ray heating rate, and models of the
Galactic synchrotron emission (Higdon 1979; Beuermann
Kanbach and Berkhuijsen 1985) suggest that the
cosmic-ray energy density initially declines very rapidly
(scale-height $\sim0.3$~kpc) as one moves away from
the plane of the disk, changing to a more gradual
decline further out. Thus the discrepancy suggested
by Lawrence might not be all that serious. Clearly
problematic, though, in the context of the present model,
is the high bolometric flux of the SCUBA sources. For
a 2~mJy source density of $10^3\;{\rm deg^{-2}}$
-- as per Lawrence (2000) -- assuming Euclidean source
counts, the brightest source on the sky should be
of order 240~Jy. These are continuum fluxes at
850~$\mu$m ($3.5\times10^{11}$~Hz), so the bolometric
flux of the brightest source should be of order
$10^{-9}\;{\rm erg\,cm^{-2}\,s^{-1}}$. The
present model predicts a maximum bolometric flux
(\S2.1) which is only 1\% of this value, based on
cosmic-ray heating. This conclusion is largely
independent of the spectrum of the emission, within
the plausible range of possibilities. However, it is
important to note that Lawrence's analysis excluded
grey-body spectra for the SCUBA sources, preferring
a ``dusty'' thermal emission spectrum (see next
section). Lawrence's interpretation of the
SCUBA data can be tested by surveying a large area
of the sky, in order to measure the bright source
counts, and following up with detailed studies of
the brightest (presumably the closest) sources.
The Planck
satellite\footnote{{\tt http://sci.esa.int/home/planck}}
should be able to detect the closest of the individual
dense gas clouds (Sciama 2000a).

\section{Discussion}
The calculations presented in \S4.4 assume that the cold
gas emits as a grey-body. This is plausible, but not
certain. Lawrence (2001) has suggested that the
Blank Field SCUBA sources are cold, dense gas clouds,
with properties similar to those which we have considered
in this paper, and he uses data on these sources to
{\it exclude\/} grey-body spectra, preferring ``dusty''
spectra which exhibit much less flux at frequencies below
the thermal peak. In addition to this uncertainty, the
atmospheric temperature could be slightly higher than the
4.2~K we have adopted, again diminishing the predicted
power at low frequencies. We have quantified these
uncertainties by computing the mini-halo anisotropies
for two models representing the plausible range of emitted
spectra: a 4.2~K grey-body spectrum (the reference model
used in the foregoing sections of the present paper),
and a 4.9~K ``dusty'' spectrum (i.e. $I_\nu\propto\nu^4$
at low frequencies). The results are shown in figure 5 for
the $l\simeq50$ peak of the model anisotropies, together
with models (adapted from Tegmark and Efstathiou 1996)
for the known Galactic foregrounds (diffusely distributed
dust, synchrotron and free-free emission). The plot is
made for $|b|=90^\circ$, and at lower latitudes all of
the Galactic contributions scale up as $1/|\sin b|$.

From this figure we can see that, in the case of a
grey-body spectrum, mini-halo anisotropies are predicted
to be the dominant Galactic foreground for observing
frequencies in the range 40--80~GHz, at multipoles
$l\sim50$. Furthermore, the predicted power contribution
amounts to $\sim\,$5\% of the cosmic background
anisotropies, at $|b|=30^\circ$ and may therefore
be detectable with data from the MAP satellite. On
the other hand, if the emission spectrum of the clouds
is ``dusty'' we can see that the predicted mini-halo
anisotropies are negligibly small at all frequencies,
and could not expect to be detected by any current
or planned instrument.

The foregoing considerations are specific to the
anisotropy power spectrum, but it may be possible
to detect microwave emission from the mini-halos
in other ways. In particular we note that the brightest
mini-halos may be detectable as individual sources.
In this case those sources should also be bright in
gamma-rays, and a sensible strategy is therefore to
search for microwave counterparts to the Unidentified
EGRET sources (Walker, Mori and Ohishi 2003). We also
note the possibility of detecting the predicted microwave
foreground component via its mean intensity which, as
noted earlier, is much larger than the anisotropy level.
For grey-body emission the mean intensity is roughly
$30/|\sin\,b|\;{\rm\mu K}$ in the Rayleigh-Jeans
regime.

The fact that the present model naturally yields a
degree-scale anisotropy peak, in the form of
low-temperature thermal radiation, suggests that it
might be possible to construct a model in which the
observed microwave anisotropies are interpreted
entirely in terms of emission from mini-halos. In
support of this idea we note that (i) the frequency
dependence of the anisotropies would be quite small
(assuming a grey-body spectrum at 4.2~K -- see figure 5),
and (ii) the heating rate of the clouds might be
rather larger than our estimate, either because of
a very large population of low-energy cosmic-rays
(see the discussion in \S4.1), or because there are
other heat sources which are more important than
cosmic-rays. However, a generic prediction of any
such model is a strong latitude dependence of the
peak power, and this is at odds with existing
constraints on any latitude dependence (Griffiths
and Lineweaver 2003).

\section{Conclusions}
We have calculated the microwave anisotropies which are
expected in a model where the dark halo of the Galaxy is
composed entirely of cold, dense gas clouds, heated
by cosmic-rays. There are contributions to the anisotropy
arising from the Poisson noise of individual clouds, and
from the Poisson noise of clusters of clouds. The latter
contribution has been computed under the assumption of a
CDM-like mass-spectrum for the clustering, and we find that
it peaks at $l\sim50$. The Poisson noise contributed by
individual clouds is, for $l\ll3\times10^8$, isotropic and
sufficiently small that it will be extremely difficult to
detect. A better approach to constraining this component
is to attempt to detect the individual clouds themselves,
and this is best done with high-resolution all-sky surveys
at high frequencies. The Planck satellite will collect
data of this type and should detect $\ga10$ such sources.

The predicted power due to clusters of clouds increases toward
the Galactic plane as $1/\sin^2b$. If the clouds emit as
grey-bodies, then such clusters are predicted to be the
dominant Galactic foreground over the range 40--80~GHz,
at multipoles $l\sim50$. For these circumstances the
predicted power contribution is $\sim\,$5\% at $|b|=30^\circ$,
and by virtue of its latitude dependence this component
may be detectable in MAP data. On the other hand, if
the clouds have a ``dusty'' emission spectrum, then their
contribution to the anisotropies is too small to be
detected, being swamped by other Galactic foregrounds
at all frequencies.

\acknowledgements 
We thank Mark Wardle for providing his radial temperature
profile for a polytrope with index ${\rm n}=3/2$, and
Lister Staveley-Smith for a critical reading of the manuscript.

\end{document}